# The Planetary Virtual Observatory and Laboratory (PVOL) and its integration into the Virtual European Solar and Planetary Access (VESPA)

R. Hueso[1], J. Juaristi[1], J. Legarreta[2], A. Sánchez-Lavega[1], J. F. Rojas[1], S. Erard[3], B. Cecconi[3], Pierre Le Sidaner[4]

[1] Dpto. Física Aplicada I, Escuela de Ingeniería de Bilbao, UPV/EHU, 48013, Bilbao, Spain.

[2] Sistemen Ingeniaritza eta Automatika Saila, Bilboko Ingeniaritza Eskola UPV/EHU, 48013, Bilbao, Spain.

[3] LESIA, Observatoire de Paris, CNRS, PSL Research University, UMPC, Université Denis Diderot, 5 Place Jules Janssen, Meudon, France.

[4] DIO, Observartoire de Paris, CNRS, PSL Research University, 61 Av. de l'Observatoire, Paris, France.

___________________________________________

*Corresponding author:* Ricardo Hueso

Full Address:  Dpt. Física Aplicada I
            Escuela de Ingeniería de Bilbao
            Plaza Ingeniero Torres Quevedo, 1 48013 Bilbao, Spain

Tel: 00 34 94601 4262
Fax: 00 34 94601 4178

E-mail: ricardo.hueso@ehu.eus

**Abstract**

Since 2003 the Planetary Virtual Observatory and Laboratory (PVOL) has been storing and serving publicly through its web site a large database of amateur observations of the Giant Planets (Hueso et al., 2010a). These images are used for scientific research of the atmospheric dynamics and cloud structure on these planets and constitute a powerful resource to address time variable phenomena in their atmospheres. Advances over the last decade in observation techniques, and a wider recognition by professional astronomers of the quality of amateur observations, have resulted in the need to upgrade this database. We here present major advances in the PVOL database, which has evolved into a full virtual planetary observatory encompassing also observations of Mercury, Venus, Mars, the Moon and the Galilean satellites. Besides the new objects, the images can be tagged and the database allows simple and complex searches over the data. The new web service: PVOL2 is available online in http://pvol2.ehu.eus/, contains a fully functional search engine and constitutes one of the many services included in VESPA (Virtual Europan Solar and Planetary Access). Data from PVOL2 can also be served from the VESPA portal using the EPN-TAP protocol. PVOL2 also provides long-term storage for amateur observations containing about 30,000 amateur images starting in the year 2000. Current and past observations from the amateur community provide a global view of the Solar System planets over the years with several possibilities for scientific analysis and amateur astronomers involvement in planetary science.

# 1. Introduction

The potential of amateur astronomers to provide valuable data in the study of the Solar system has long been acknowledged. The first request to amateur astronomers to observe solar system bodies comes probably from the times of Edmund Halley and his call to observe the 1715 total eclipse of the Sun which crossed England with the aim to determine the parallax of the Sun and the Moon from multiple observations (Halley, 1716; Marshall et al., 2015). The space era, with spacecraft visits to the planets, has motivated the need for timely monitoring of the planets that provide the context to the detailed observations acquired from robotic missions. Over the last two decades, continuous progress in image acquisition from low-cost equipment have resulted in a revolution on amateur astronomy that has largely impacted the field of planetary sciences (Mousis et al., 2014) and has resulted in several professional and amateur collaborations (PRO-AM in the following). Three key elements provide a successful PRO-AM collaboration: (1) Amateurs have time availability and flexibility to react rapidly to interesting activity in the planets; (2) They hold a large expertise in several fields including high-resolution imaging in the visible and are even starting to provide spectroscopic data of scientific interest; (3) They can observe the planets close to the Sun, at angular distances that are often prohibited to large telescopes and they can also observe at very low elevation angles that in many cases are not available to professional telescopes. However, data gathered by amateur astronomers need to be communicated to researchers and the public and stored for analysis of the temporal changes of the planets atmospheres and other phenomena. The accumulation of data by many different amateurs in an online-database of observations constitutes a virtual global observatory that allows a research centred in the temporal evolution of particular phenomena.

Planetary images obtained by amateurs constitute a long-term, at times dense, survey of the surfaces (Mercury and Mars) and atmospheric activity of the planets (Mars to Neptune). When a planet approaches opposition and is at a small angular separation from the Sun, amateur astronomers distributed throughout the Earth can provide almost continuous images when large optical telescopes cannot observe the planet for security reasons. The problem of avoiding unsafe proximity to the sun is particularly important for the inner planets Venus and Mercury. It is also important in the study of atmospheric phenomena on Jupiter, which develops frequently and sometimes near solar conjunction. Finally, the quality of the observations is in some cases outstanding. Many observers acquire regularly diffraction limited images of the brightest planets. In some cases, especially if features have an intrinsic high contrast, spatial scales beyond the diffraction limit of the telescopes can be resolved resulting in super resolution images. This works because low levels of atmospheric turbulence shift acquisition frames by a few to fractions of pixels (Marsh et al., 2004) allowing to acquire super resolution images after processing a large stack of frames which contains a very high signal to noise ratio and using a variety of processing tools (wavelets transforms, high-pass filters, deconvolution). A typical example is amateur observations of Saturn rings, where the Encke division (330 km) is observed over most of the ring system with telescopes with apertures of only 40 cm (limited by diffraction to resolutions of 2,300 km or 0.3'' observing at visible wavelength of 500 nm). Several references on amateur high-resolution observations are available in the amateur literature (Grafton, 2003; Buick 2006; Pellier et al., 2015) and also in Mousis et al. (2014).

A first major effort to gather coordinated observations from professional and amateur astronomers took place during the Earth approach and perihelion of comet Halley in 1986. This coordinated campaign was called The International Halley Watch IHW (Newburn, 1983) and was endorsed by the International Astronomical Union in 1982 producing a large quantity of data (Sekanina and Fry, 1991) that is currently available at NASA's Planetary Data System (PDS)[1]. In a similar way to the IHW, and midway between the flybys of the Pioneers and Voyagers to explore Jupiter and outer planets and the Galileo mission, The International Jupiter Watch (IJW) was created to coordinate the study of time variable phenomena in the Jovian system (Belton et al., 1989; Russell et al, 1990). This group developed strong links with the amateur community at the time of the Galileo mission to Jupiter organizing a web site repository of amateur observations. That initial repository of Jupiter images was first hosted at the New Mexico State University under the leadership of R. Beebe. This organization expanded later to cover time domain phenomena on the Outer planets forming the International Outer Planets Watch (IOPW) which covers different disciplines from Atmospheres of the Giant planets to Titan and auroras and magnetospheres[2]. The image repository containing amateur observations of the atmospheres of the Giant Planets was moved later to the Universidad del País Vasco in Spain in 2003 where it evolved into the first version of PVOL (Hueso et al., 2010a).

---

[1] http://pdssbn.astro.umd.edu/data_sb/missions/ihw/index.shtml
[2] http://www-ssc.igpp.ucla.edu/IJW/

In parallel with those efforts, and following the introduction of digital imaging in astronomy along the 80s, amateur organizations created their own repositories of images organized by different planets and with data ordered in time. Special efforts were done in the case of Mars (e.g. Troiani et al., 1996) and Jupiter as mentioned above. Pro-Am collaborations in the field of comets also count with image repositories of observations with very different needs to planetary images. The most popular of these repositories of amateur images are listed in Table 1.

**Table 1: Most prominent image repositories of amateur images**

| Acronym / Name | Full name | Web site |
| --- | --- | --- |
| ALPO - Japan | Association of Lunar and Planetary Observers - Japan | http://alpo-j.asahikawa-med.ac.jp/ |
| ALPO* | Association of Lunar and Planetary Observers - USA | http://alpo-astronomy.org/ |
| SAF | Société Astronomique de France (SAF) | http://www.astrosurf.com/planetessaf/ |
| UAI | Unione Astrofili Italiani (UAI) | http://pianeti.uai.it/index.php |
| COBS** | Comet Observation Database | http://www.cobs.si/ |

(*) ALPO contains several sections including comets, meteors and Moon images. The data amount in different sections varies largely. (**) The COBS database contains also analysis of the observations including light curves of many comets based on observations by many different amateur contributors.

Very recently, the Rosetta mission to comet 67P/Churyumov-Gerasimenko has also launched an important pro-am collaborative effort mainly distributed through Facebook groups (Yanamandra-Fisher, 2016). Currently, there are two active planetary space missions requesting the support of amateur observers: The Juno mission to Jupiter (Hansen et al., 2014) and the Akatsuki mission to Venus (Nakamura et al., 2016). The Juno mission hosts a very important collaboration with amateur astronomers to provide image context to the JunoCam observations and contains its own database of Jupiter amateur observations[3]. The Akatsuki mission has created its own webpage for amateur contributions[4]. Other previous space missions, such as Venus Express, have also requested support from amateurs developing their own database of amateur Venus observations (Barentsen and Koschny, 2008). The data in those PRO-AM projects is vital during the development of the missions but may become forgotten later on due to the dispersion of amateur data between the ensemble of image repositories, local and global databases, specialized blogs (an example is the Ice in Space blog in Australia[5]), personal pages and facebook astronomical groups.

---
[3] https://www.missionjuno.swri.edu/

Here we present the second version of the Planetary Virtual Observatory and Laboratory (PVOL2). This is a database of amateur observations of Solar System planets and the Moon that grew from PRO-AM collaborations in the field of the Giant Planets and the previous database of the International Outer Planets Watch (Hueso et al., 2010a). PVOL2 inherits the previous amateur observations of the Giant Planets (years 2000-2016) and adds images of Venus and Mars with demonstrated scientific potential (Sánchez-Lavega et al., 2016b; Parker et al., 1999; Sánchez-Lavega et al., 2015) as well as images of Mercury, the Moon and the Galilean satellites. PVOL is one of the services of the Virtual European Space and Planetary Access infrastructure (VESPA) and can be consulted from the VESPA search interfaces through the use of the EPN-TAP protocol (Europlanet Table Access Protocol; Erard et al., 2014). VESPA itself is described in Erard et al., (2017). The aim of the PVOL2 service is to store the increasing number of image files obtained by the large community of amateur astronomers providing a friendly but powerful search capability of the observations, a link between professional astronomers and the amateur community, and a long-term plan to curate the data and make them available years after they were acquired.

## 2. PVOL2: Technical description

PVOL2 is a web service inspired in Virtual Observatory concepts (Szalay and Gray, 2001; Graham et al., 2008) hosting images of Solar System planets together with ephemeris information and appropriate labels in the most modern images. PVOL2 is hosted in an Apache 2.4 web server with an Apache Tomcat 7 servlet engine and is available as a web service on http://pvol2.ehu.eus. The previous data in the original PVOL server has been migrated from MySQL to PostgreSQL (both popular database management systems where SQL means Structured Query Language; PostgreSQL is free open source and is the development platform recommended in the framework of the VESPA project) and the database structure has been improved adding other fields that are intended to give more information about the data stored in PVOL2. In PVOL2 a single image file may contain multiple observations of the same planet in different filters and time steps and constitutes different elements in the database, all of them searchable separately.

---

[4] https://akatsuki.matsue-ct.jp/
[5] http://www.iceinspace.com.au/

Ephemerides relevant to each observation are added to each image or data registry following celestial mechanics algorithms from Meeus (1991) and Archinal et al. (2011). Ephemerides are precomputed and stored in time tables for each planet and year so that when a new observation becomes available its ephemeris values are interpolated from the closest two time steps in the precomputed files. Ephemerides include the equatorial size of the planet in arc sec, the phase angle of the observation (especially important for Venus and Moon observations) and the central meridian of the planet in different rotation systems defined by the International Astronomical Union (IAU). For instance: Jupiter longitudes can be defined in system I (useful to study equatorial latitudes), system II (useful for mid latitudes) and system III (associated to the internal magnetic field and the assumed rotation system of the planet). Other planets incorporate different systems of longitudes according to their definition by the IAU. The sole exception is Venus where a system II (not defined by the IAU but useful for the study of the upper cloud layer of the planet) is provided and defined with a rotation period of 4.2 days close to the atmospheric superrotation period at the upper cloud layer and equivalent to an angular velocity of -85.71429°/day (Pellier et al., 2015). Table 2 summarizes the available systems of longitudes for each planet.

**Table 2: Longitude systems defined for each object in PVOL2.**

| Object | System I | System II | System III | Reference |
|---|---|---|---|---|
| Mercury | Surface | --- | --- | [1] |
| Venus | Surface | Upper Clouds | --- | [1, 2] |
| Mars | Surface | --- | --- | [1] |
| Jupiter | Eq. latitudes | Mid latitudes | Interior | [1] |
| Saturn | Eq. latitudes | --- | Interior (*) | [1] |
| Uranus | --- | --- | Interior | [1] |
| Neptune | --- | --- | Interior | [1] |
| Moon | Surface | --- | --- | [1] |
| I Io | Surface | --- | --- | [1] |
| II Europa | Surface | --- | --- | [1] |
| III Ganymede | Surface | --- | --- | [1] |
| IV Callisto | Surface | --- | --- | [1] |

[1] Archinal et al. (2011). [2] Pellier et al. (2015). (*) The interior rotation of Saturn is only approximate and its precise values is still debated (Sánchez-Lavega, 2005). The interior rotation is assumed to be the one derived at the time of the Voyager missions and we use the value recommended in Archinal et al. (2011).

Figure 1 summarizes the technical description of the PVOL2 service.

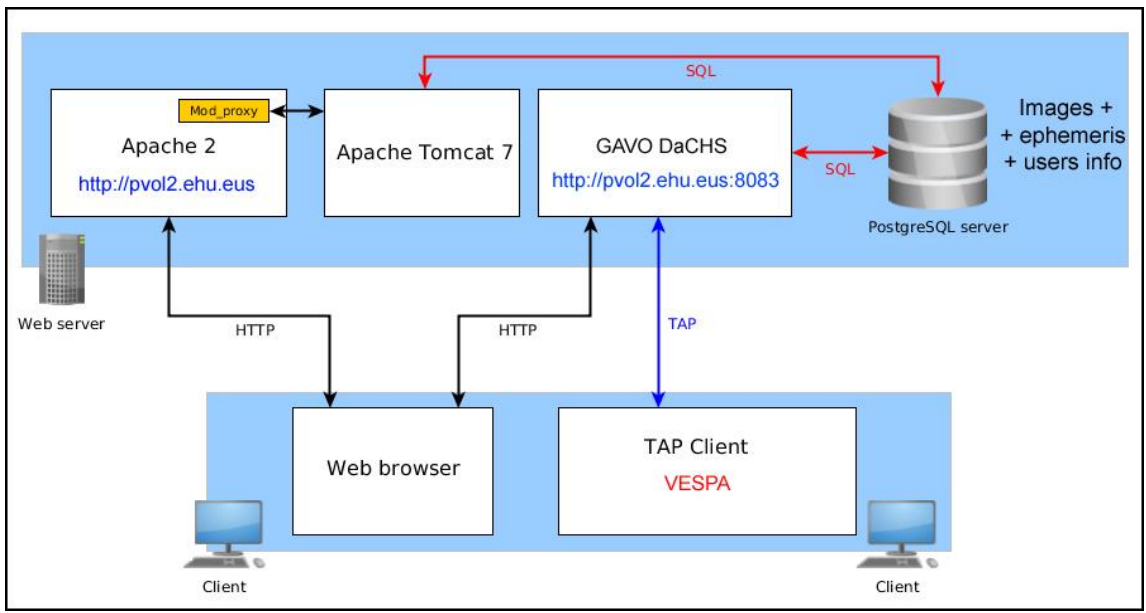

**Figure 1: Summary of the PVOL2 service.** The service can be consulted on a web browser on http://pvol2.ehu.eus or from the VESPA portal, or from tools related with Virtual Observatory projects like CASSIS, 3Dview, or TAP (Table Access Protocol) clients.

2.1. Data and users management

Users can submit their images via e-mail to: pvol@ehu.eus. They can also register in the system and get a password protected access to upload their own images. Whether the system manager or an outside user uploads the data, this operation is performed using a web form that automatically fills most of the content of the different fields describing the data from a filename following a given convention (Figure 2). If the files are provided with other name conventions the user has to fill-in the upload form. The image file stored in PVOL is stored with a name that follows the convention in Figure 2. Image files may contain one or several observations of the same object in different filters with only the first one affecting the file name convention. Cylindrical projections in different longitude systems, North and South polar projections and movies can also be uploaded in the database.

**Object Date _ Time _ filter _ observer.fileformat**

**Examples:**

| | |
|---|---|
| *j2016-05-03_11-09_ch4_cg*.jpg | Jupiter image acquired on 2016-05-03 at 11:09 UT using a methane filter by observer cg (Cristopher Go) |
| *s2016-10-23_09-51-54_r_tba*.png | Saturn image acquired on 2016-10-23 at 09:51:53 UT using a red filter by observer tba (Trevor Barry) |
| *s2016-10-17_22-01_rgb_pe*.jpg | Saturn image acquired on 2016-10-17 at 22:01UT using rgb filters (color image) by observer pe (Peter Edwards) |
| *Mars2016-10-25_16-05_ir685_cfo*.jpg | Mars image acquired on 2016-10-25 at 16:05 UT using an IR filter at 685nm by observer cfo (Clyde Forrester) |
| *v2016-10-27_07-39_uv_pmi*.png | Venus image acquired on 2016-10-27 at 07:39 UT with an UV filter by observer pmi (Phil Miles) |

**Figure 2: File names conventions on PVOL.** Examples of file names in PVOL and their meaning.

Since it is almost impossible to request to a wide community of very different users to stick to the same name criteria, we have developed external programs that translate common name files used by individual observers in the PVOL name format summarized in Figure 2. This allows uploading large numbers of files provided by common contributors quickly.

PVOL2 incorporates management tools for users and the data. Each image file and data registry is associated to an individual user. Individual users can be given an individual password that allows him to upload his own data and modify it later in case that errors are detected. Management tools to add new users or delete user privileges are also incorporated online.

2.2. Searches over the database

The PVOL2 service hosts a search engine designed using a web framework with internal processing in Java. Observations from all observers, or just from a single observer, can be searched with a combination of conditions including desired dates. Searches for images containing a central meridian between a given range of longitudes in a particular longitude system are easily produced by the database allowing for instance searches of specific regions on the surface of Mars or searches of particular features on Jupiter's atmosphere close to a given central meridian. Moon images are tagged so that searches of specific craters are possible by filling a "Feature" field in the search engine. A particularly useful capability to look for atmospheric features that drift in longitude over the planet with a given drift rate (degrees/day) has been incorporated as a tool for atmospheric studies of the giant planets. This can be done in any of the three available systems of longitudes for Jupiter. The database blocks the non possible options (for instance only system I is available for Mars). Images, movies, or maps of different types including maps in different longitude systems can be searched separately or combined. All fields that are filled by the user in the search form are combined to provide a list of observations.

Figure 3 shows the web search form using two examples of different searches. The first example results in all observations of Jupiter since the initial date in the form. The second example shows only Jupiter images between the given dates obtained by a single amateur astronomer identified by his name (Cristopher Go in the example) in a given image filter (ch4, which is the code for amateur filters sampling the strong methane absorption band at 890 nm) that have the central meridian of the planet close to 300º using System III longitudes. The lower part of the search form allows to show only images, movies, maps or specific projections. The default is that all data types are searched.

Results can be retrieved as galleries of files (paginated and shown with 50 images per page, as a default option, or paginated with 100 or 200 images per page) or as a list of files. Examples of the results from searches on Figure 3 are shown in Figure 4. Collections of images can be downloaded using a single zip file that is created from the query when clicking the link to the zip file in the results page. A limit of 100 images is imposed over this zip file. Individual images are always available as hard-links. For instance: The first image in the search shown in Figures 3B and 4B is always available as http://pvol2.ehu.eus/pvolimages/jupiter/j2016-05-03_11-09_ch4_cg.jpg

**Figure 3: Example searches on PVOL2.** (A) Simple search of Jupiter images from a given date. (B) Advanced search of Jupiter images in a time window obtained by a single observer in a particular filter and with constrains on the planet's central meridian. The lower part of the form allows including or excluding different data types in the search engine.

Even more sophisticated searches, like finding images that have a central meridian that drifts in longitude with a given angular velocity or drift rate (in deg/day) from a reference date and longitude System are possible and useful when looking for faint features in Uranus and Neptune or for fast drifting features in Jupiter. This type of search is documented on the PVOL web site. See:
http://pvol2.ehu.eus/pvol2/search/driftrate.

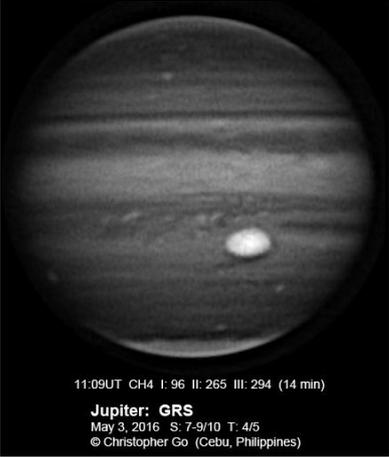

**Figure 4: Results from the searches identified in Figure 3.** (A) List of a large collection of files. (B) Gallery of images (collection of images that can be viewed with the web browser). Only the first of six images are shown in this figure.

## 2.3. PVOL2 searches and access through VESPA

PVOL2 can also be consulted from the VESPA query portal (http://vespa.obspm.fr) which performs global queries over all data services supporting the EPN-TAP (Erard et al., 2014) protocol. The PVOL EPN-TAP service is located at http://pvol2.ehu.eus:8083 and relies on a GAVO DaCHS (German Astrophysical Virtual Observatory Data Center Helper Suite) server. Although the VESPA user search interface is optimized for EPN-TAP, the service is responsive to queries issued from any tool using the Table Access Protocol (TAP), such as TOPCAT (Tools for Operation on Catalogs and Tables) (Taylor, 2005) or TapHandle, which support more general queries that can be launched at: http://pvol2.ehu.eus:8083/tap. Direct SQL queries using the Astronomical Data Query Language (ADQL) can also be formulated via web on the DaCHS client in the address http://pvol2.ehu.eus:8083/adql. Results consist in a list of images matching the query with their description. Answers to EPN-TAP queries always include a complete description of each image, a thumbnail, and the access URL; visual selection can be performed from the VESPA user interface on the basis of the thumbnail. Answers to lower-level TAP queries only include the requested parameters. Examples of web queries using ADQL are provided in table 3.

Table 3: ADQL queries in the EPN-TAP service on PVOL2

| Query | Description |
| --- | --- |
| select * from pvol.epn_core | Select all the observations |
| select * from pvol.epn_core where observer_code = 'cg' | Select the observations submitted by user 'cg' (Christopher Go). |
| select * from pvol.epn_core where target_class = 'planet' | Select all the planet observations |
| Select top 10 * from pvol.epn_core | Select the first 10 observations |
| select * from pvol.epn_core where creation_date > X and creation_date < Y | Select the observations created between dates X and Y, in Julian days |

Web queries return results in different formats including HTML, VOTable and CSV files. The recommended VO-access to PVOL is from the VESPA portal, which is designed to be more intuitive and less technical. VESPA can query all EPN-TAP services together and will therefore also provide results from similar services such as BDIP[6] (Base de Données d'Images Planétaires; historical images of planets in the 20th century), allowing for longer-term variations studies An example of a query in VESPA returning data from BDIP is shown in Figure 5. VESPA also directly connects to imaging tools and support for planetary coordinate systems in Aladin is being studied. However the VESPA portal is not user friendly for amateur astronomers and the PVOL2 service is the recommended access gate to the data for amateur astronomers and the general public.

---

[6] http://www.lesia.obspm.fr/BDIP/

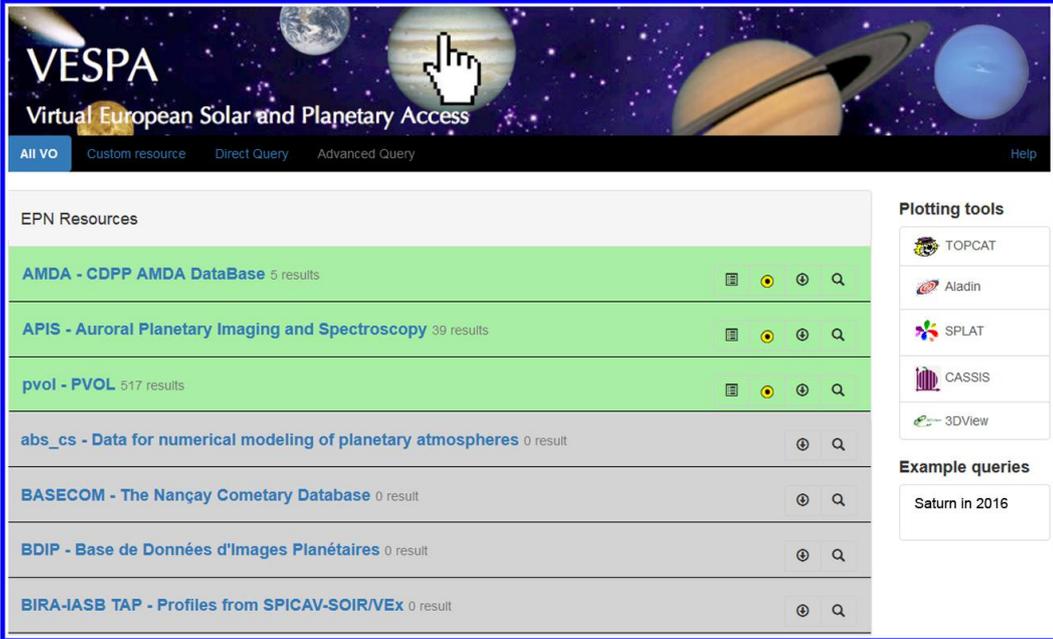

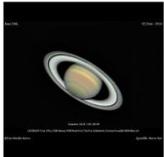

**Figure 5: Example of a VESPA query for images of Saturn on a given time range.** (A) VESPA results from different services appear in green. (B) Results of the query through PVOL. Each result also shows a thumbnail of the image and a direct link to the data in the access_url field.

## 2.4. Data Management plan

Backups of the data are programmed weekly and contain two parts. On the one hand, the full set of images stored in different directories (one for each planet, one for satellites and one for the Moon) are compressed into a single file, respecting the folder structure and adding a time stamp to the file name so that several zip files generated by users are stored temporarily to serve nearly simultaneous queries. On the other hand, the database, where all the information regarding PVOL (users, observations) is stored, is weekly dumped to an SQL file also with a time stamp in its name. Additionally, the database used for external queries using EPN-TAP is also dumped into a different SQL file with information on date and time. The database can be restored from the SQL file using tools in PostgreSQL and the images can be restored by simply decompressing the backup file. An external copy of the backup files and software are copied and stored in an outside driver in a different office of the PVOL2 server.

## 3. Data in PVOL2

### 3.1. Current data

PVOL2 contains all the image files from the previous PVOL version of the database at the time when it only contained images of the Giant Planets. Images of Jupiter are available since the year 2000 and a few professional observations are also incorporated in the database. Since August 2016 the database contains also regular observations of Venus and Mars. Data demonstrating scientific and archiving capabilities of observations of Mercury, the Moon and the four Jupiter Galilean satellites are also available. We regularly contact different amateur astronomers to provide permission to upload their past observations into the PVOL2 service so that the database is actively being populated backwards with past data.

In almost all cases, images in the PVOL2 database are highly processed to attain the highest spatial resolution available in the data. This means that the images constitute sharp observations of the planet surfaces and cloud systems but no photometric information that could be used to characterize the albedos of the objects or perform photometry of the cloud systems. Those unprocessed versions of the observations can be requested to individual observers if required.

Most of the observations in the database are color-composited RGB images obtained in broad-band filters and sent in different image formats: jpg, png, tiff and gif. Some sets of observations are uploaded as movies in formats gif, avi and mp4 but only gif files will be visualized in the PVOL2 web site and all other formats must be downloaded for visualization. PVOL2 hosts almost 30,000 image files with about 900 planetary maps (mainly Jupiter cylindrical maps and polar projections of Jupiter and Saturn) and 500 movies. Table 4 gives a short description of the number of observations per object. As of November 2016 the total amount of data is about 4.0 GB. Updated statistics about the data are automatically provided by the system in a link "Statistics" in the bottom part of the webpage.

**Table 4: Number of observations in PVOL2 available in January 2017**

| Planet | Total number | Data for 2016 | Period |
|---|---|---|---|
| Mercury | 3 | 0 | 2007-2011 |
| Venus | 208 | 145 | 2015-2017 |
| Mars | 208 | 177 | 2014-2017 |
| Jupiter | 22342 | 2080 | 2000-2017 |
| Saturn | 7279 | 517 | 2001-2017 |
| Uranus | 412 | 80 | 2002-2016 |
| Neptune | 230 | 96 | 2006-2016 |
| **Other datasets** | | | |
| Moon | 190 | 13 | 2004-2016 |
| Io | 3 | 0 | 2009-2014 |
| Europa | 2 | 0 | 2008-2014 |
| Ganymede | 15 | 2 | 2008-2016 |
| Callisto | 3 | 1 | 2010-2016 |

3.2. Observers and telescopes

There are about 300 observers, with some contributions coming from amateur organizations or observatories instead of individuals. Observers are distributed worldwide including observers contributing from equatorial latitudes (the best suited for planetary imaging). The location of Jupiter observers in the period 2010-2013 (roughly representing the global distribution of observers contributing to PVOL) appears in

Figure 6. They typically use 15-40 cm telescopes, different Barlow lenses to amplify the focal length of their telescopes, filter wheels and fast acquisition cameras (Mousis et al., 2014). The database also contains many planetary observations obtained at the 1.06-m telescope at the Pic du Midi observatory in France since the year 2000. This telescope runs a strong program of collaboration with amateurs. For historical reasons the database also contains a few observations of Jupiter by the InfraRed Telescope Facility (IRTF) in Hawaii.

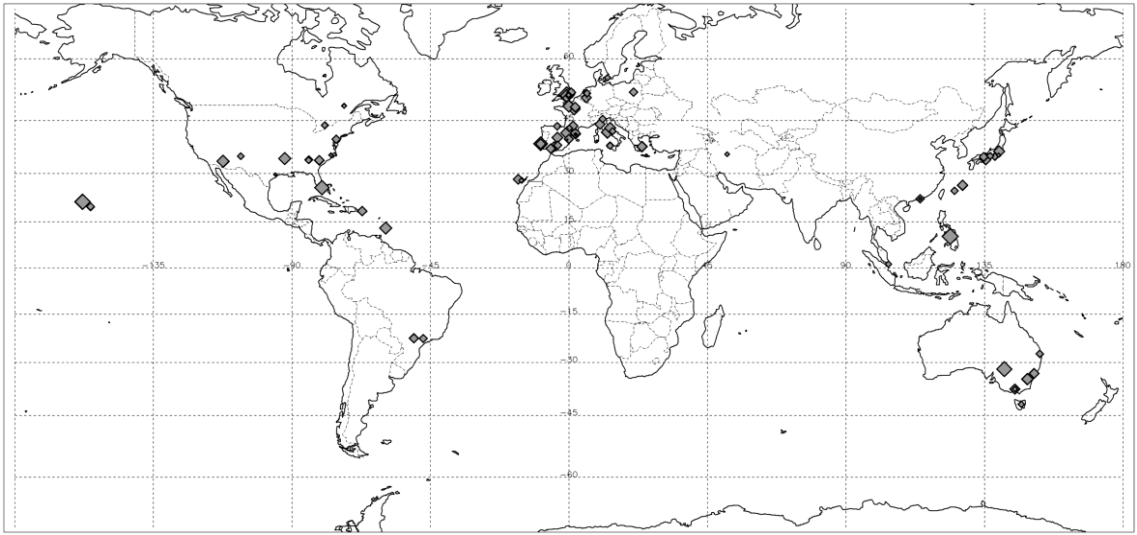

**Figure 6: Spatial distribution of Jupiter contributors to PVOL.** The map is built from the location of Jupiter observers in 2010-213. The size of each dot is proportional to the logarithm of the number of observations per observer. There is an important lack of observers in the Middle East to Central Asia to really obtain a full network of observers. Japan is sub represented since most Japanese observers contribute to ALPO-Japan only. Efforts to gather more contributors from United States, Central and South America are under way.

3.3. Data quality

In the initial PVOL description in 2010 (Hueso et al., 2010a), and in more modern references like Mousis et al. (2014), examples of high-quality observations of the planets are extensively provided. An update of modern high-quality amateur observations is presented in figures 7 to 12. Figure 7 shows the possibilities of studying dynamics of Jupiter from comparison of high-resolution observations obtained separated in time by exactly one Jovian rotation (9.93hr) and from full maps obtained by a single or several observers on different dates. Figure 8 shows the multiple possibilities offered to study seasonal changes in Saturn from observations gathered over 14 years and the continuous increase of image quality produced by amateur astronomers.

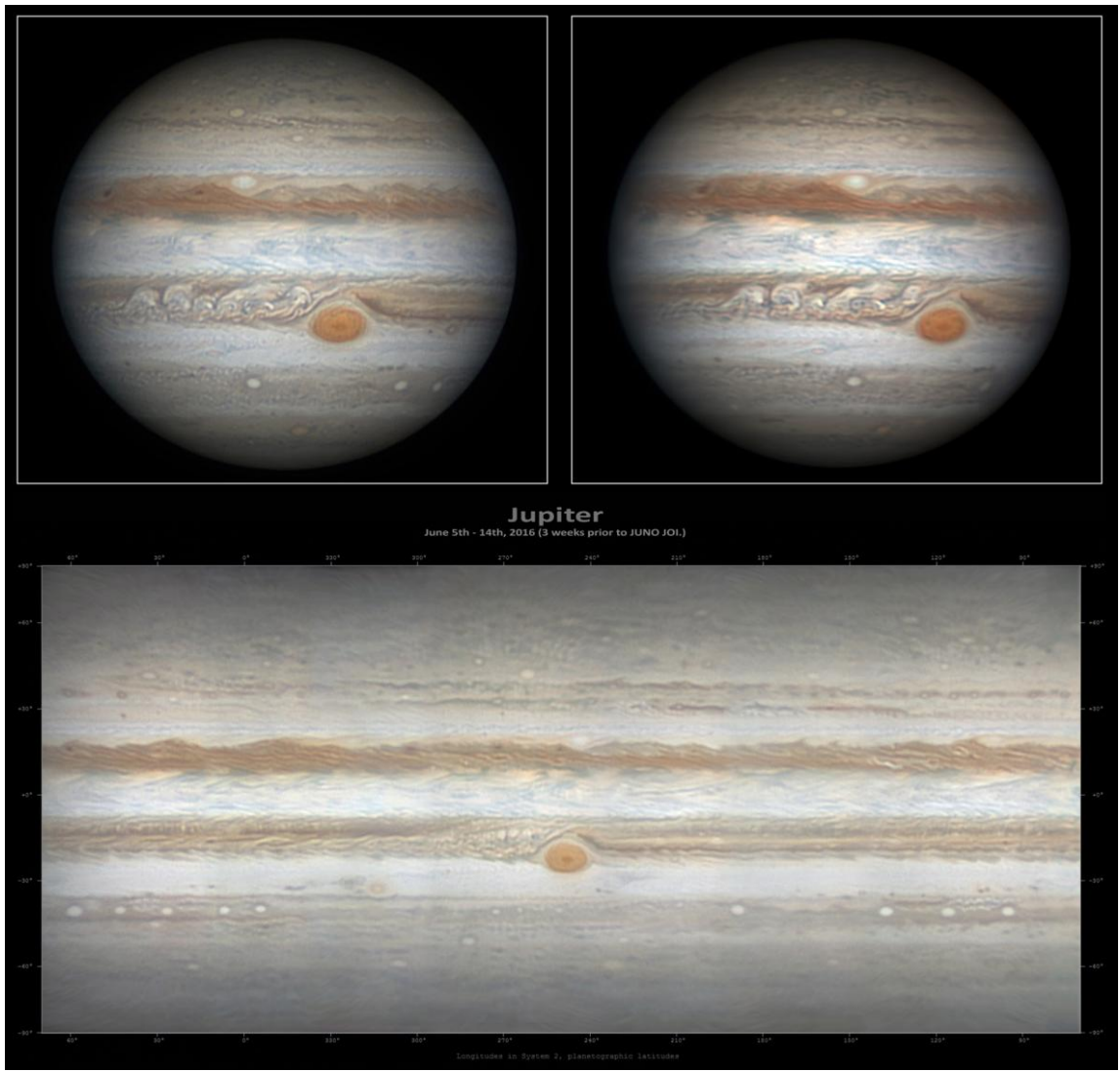

**Figure 7: Jupiter images after one planetary rotation and over a full planetary rotation.** Upper row: Left image from Damian Peach observing from Barbados on March 18, 2016 at 03:03 UT. Right image from Christopher Go observing from Phillipines one Jupiter rotation later on March 18, 2016 at 13:23. Bottom image: Cylindrical map of Jupiter processed by Damian Peach from his own observations obtained in June 5 to June 14 prior to Juno arrival to the planet. Many of these maps are computed regularly by amateur astronomer M. Vedovato and are available in the PVOL2 database and the webpage of the Unione Astrofili Italiani (UAI).

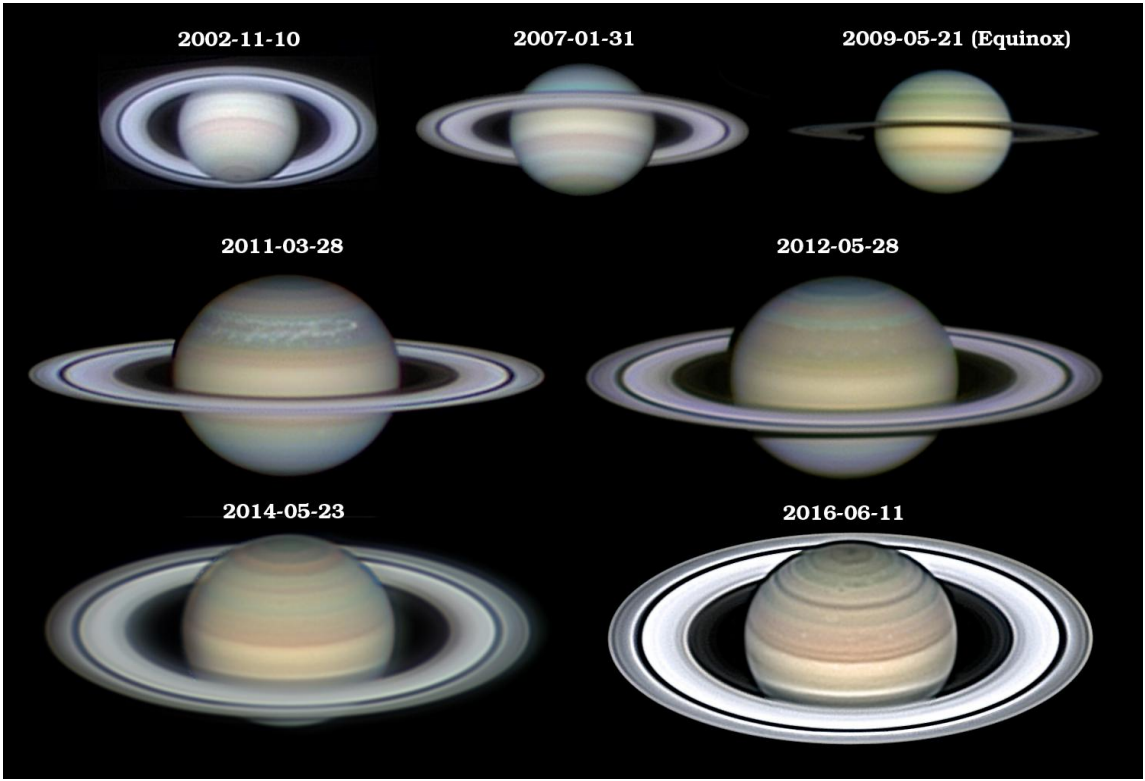

**Figure 8: Saturn images over several years.** All images from Damian Peach except for the image on 2014 that was obtained by Trevor Barry.

Two comments have to be made in the cases of Venus and Mars (not included in the previous version of PVOL), and in Uranus and Neptune, where recent important achievements have been obtained. For Venus, regular images of the upper cloud deck using ultraviolet and violet filters (typically 330-400 nm, 380-420 nm), as well as images at 1 μm sensitive to a lower altitude of the upper clouds, are now provided by several amateurs. Images of the thermal emission from the surface of the planet are also obtained by a relatively small number of observers (Figure 9). For Mars, excellent observations of the surface at long wavelengths and of the global cloud systems in short-wavelength images (blue to ultraviolet) are also available in the database (Figure 9). Amateur observations of Uranus and Neptune (Figure 10) have matured largely due to new fast inexpensive cameras with higher sensitivities at long wavelengths (>680 nm) and the use of low cost long pass imaging filters, typically starting at 610 – 700 nm and extending until the detector cutoff close to 1 μm. In recent years amateur observations have been used to provide ephemeris of atmospheric details rotating in both planets and subject of research by powerful professional telescopes. There are several active PRO-AM collaborations in Uranus (Sromovsky et al., 2012; de Pater et al., 2014) and Neptune (Hueso et al., 2016a) in the last few years.

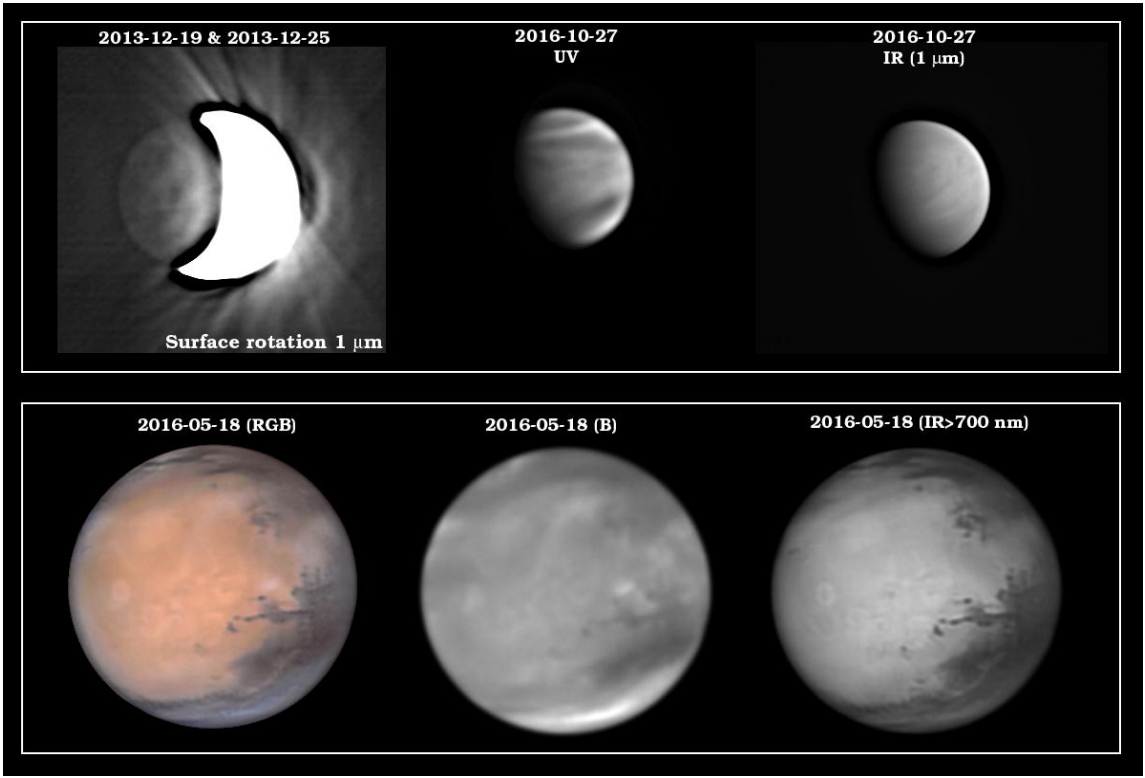

**Figure 9: Venus and Mars images.** Venus images from A. Wesley. Mars images from P. Miles.

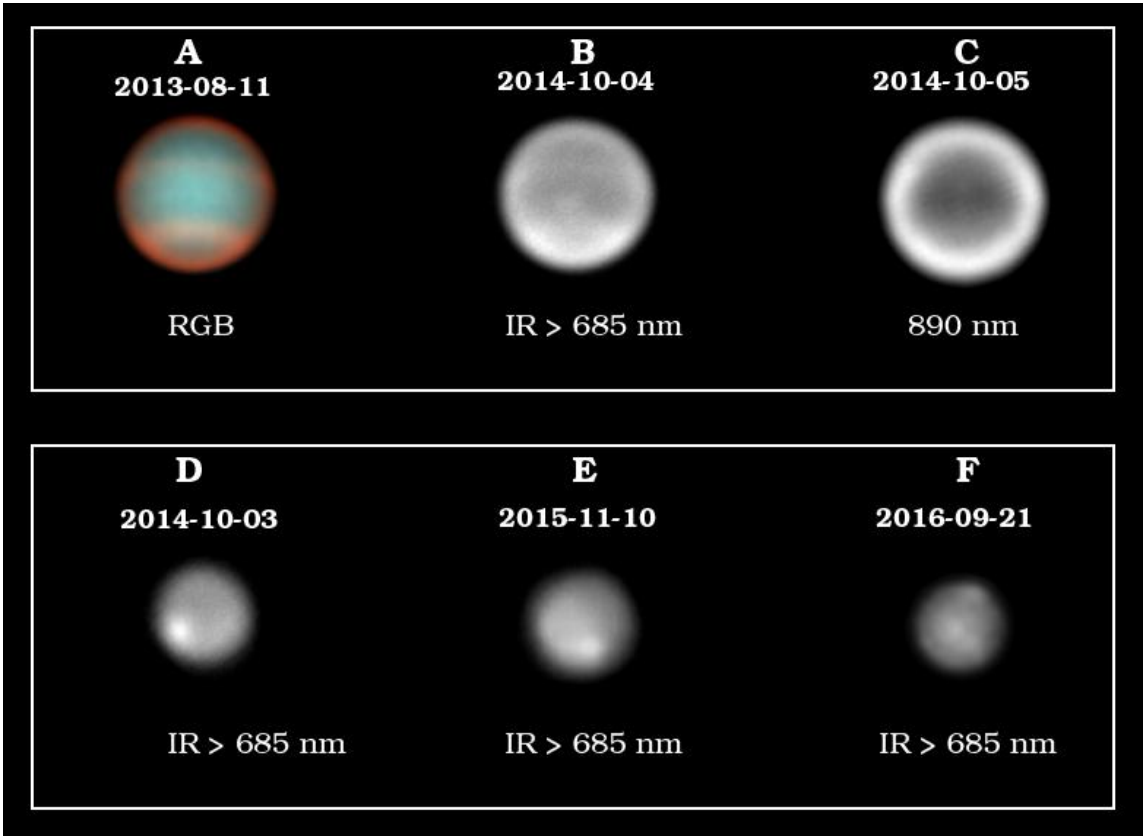

**Figure 10: Uranus and Neptune.** (A-C): Uranus images obtained at the 1.06-m telescope at the Pic du Midi observatory by amateur astronomers in collaboration with F. Colas from the Observatory of Paris. Observers are (A) J.-L. Dauvergne, S. Bouley and F. Colas. (B and C) M. Delcroix and F. Colas. (D-F): Neptune images. (D) Observation at the Pic du Midi telescope by M. Delcroix and F. Colas. (E) Observation from Phil Miles using amateur equipment. (F) Observation from D. Millika and P. Nicholas using amateur equipment.

Also, although images of the Galilean satellites by amateurs are at the limit to resolve surface features amateurs acquire such images with certain regularity. Figure 11 show examples from the PVOL2 database that range from surface features on Ganymede and Callisto to mutual encounters between Galilean satellites.

Finally, different types of images of the Moon, including eclipses and high-resolution observations, are included in the PVOL2 service attaining in some cases spatial resolutions on the order of 200 m/pixel. The Moon and satellites collection is relatively limited and will be increasing in number of observations in the near future.

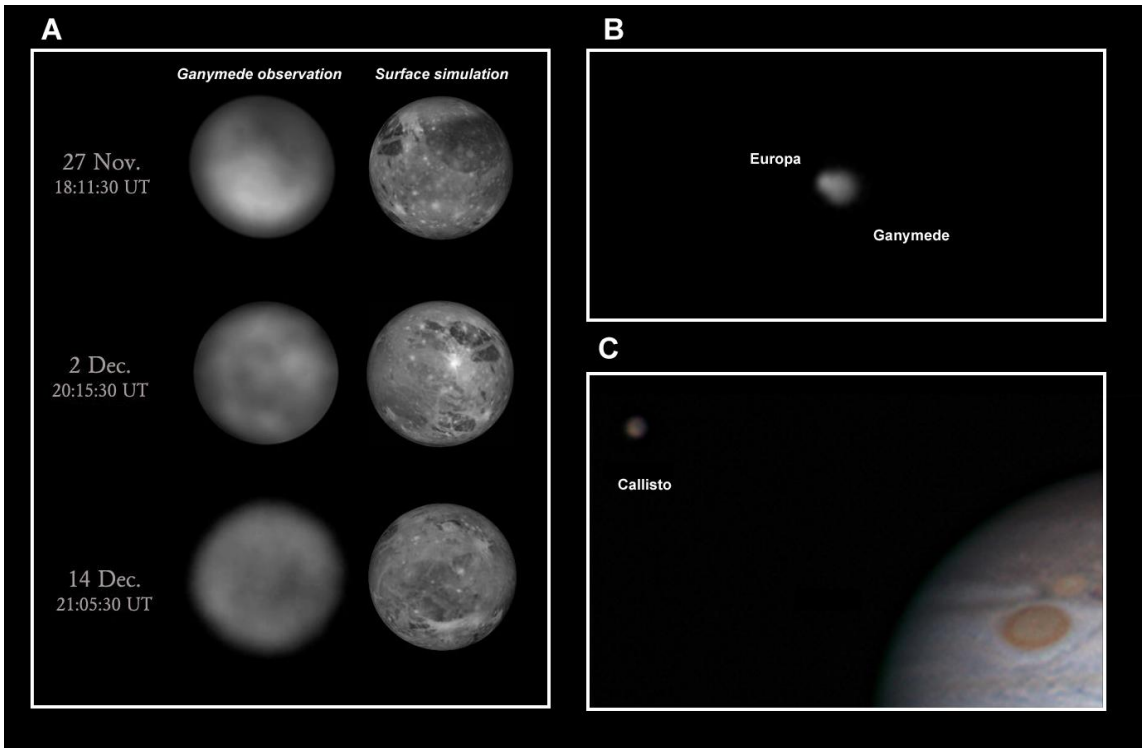

**Figure 11: Galilean satellites.** (A) Ganymede observed on three different dates in December 2011 by M. Kardasis using a 28-cm telescope in white and red light compared with the true surface of Ganymede simulated with the WinJupos software. (B) One frame of movie with a partial occultation of Ganymede by Europa on May 8, 2008 observed by A. Wesley using a 36 cm telescope. (C) Image of Callisto showing part of Jupiter obtained by E. Morales on August 2010 using a 36-cm telescope.

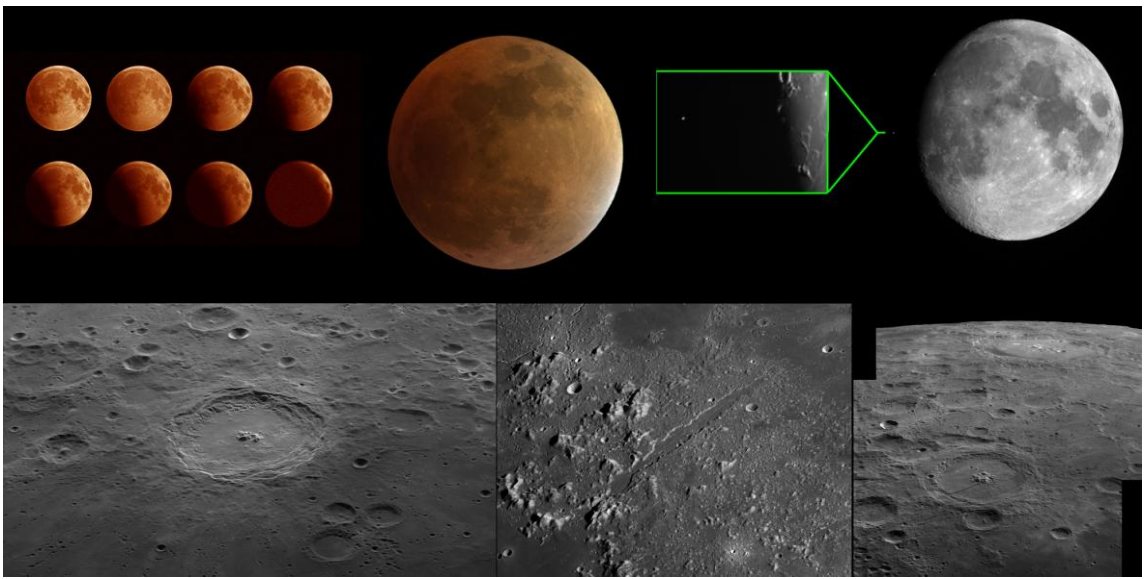

**Figure 12: Examples of lunar images on PVOL.** The Moon collection of images in PVOL is currently being expanded and contains eclipse images, stellar occultations, Moon-planet conjunctions and high-resolution observations of Moon features. The figure above contains thumbnails of images of the Moon, which sometimes can be larger than 2000 pixels in each direction with spatial resolution of particular details on the order of 200 m/pixel. Images from E. Morales (upper left and upper right images) and D. Peach (all others).

**4. Modern scientific cases for the analysis of amateur data**

In our previous paper describing the original PVOL web site (Hueso et al. 2010a) we provided a list of scientific topics that can be addressed with a long-term, fine time coverage, high-resolution images of Jupiter and Saturn. These focused on surveys of the atmospheric phenomena on the giant planets and their temporal evolution, including the onset and evolution of large to planetary-scale storms (Sánchez-Lavega et al., 2008; 2011, 2012) as well as in the characterization of impact events in Jupiter (Sánchez-Lavega et al. 2010; Hueso et al., 2010b, 2013). Since then, Mousis et al. (2014) have published an updated list of topics considering the overall observations of solar system objects by amateurs including the interest of studying Venus and Mars. Here we summarize those topics and focus on science themes not previously covered and that can be researched by the overall community of professional astronomers using data from PVOL2.

- The Moon: The database incorporates tagged images of the Moon. Images of particular craters can be found by filling the "Feature" field in the PVOL search engine. The Feature field allows auto filling so that filling a few characters is enough to find the objects of interest. Interesting locations like the Apollo 11 landing site are tagged in a few images. Conjunctions of planets like Mars, Jupiter or Saturn with the Moon can be searched in the "Feature" field as well as stellar occultations. High-resolution images of several surface features are available with some of them at spatial resolutions of 200 m/pixel or better. Images of the Moon in PVOL2 can be used for research as well as teaching and astronomy popularization activities. Current lunar images in the PVOL2 database demonstrate these possibilities.

- Satellites of the Giant Planets: The database incorporates images of the Galilean satellites of Jupiter and images of satellites mutual phenomena (transits, occultations, eclipses, all of them generally known as Phemus). The scientific use of amateur observations of Jupiter satellites mutual phenomena is well documented (Arlot et al., 1992, 2014). High-resolution observations of Ganymede may become of interest after the launch of the Jupiter ICy moons Explorer (JUICE) mission in 2022 (Grasset et al., 2013).

- Mercury: Mercury is a difficult target for most telescopes due to its small angular size and small separation from the Sun. Amateur observations of Mercury have been gathered at the time of its transits across the Sun and recent and coming space missions to the planet (Messenger and Bepi-Colombo) may motivate more amateur observations of the planet. However, scientific results from those observations are not yet demonstrated.

- Venus: Besides the overall characterization of the meteorological patterns in the planet including short-term global changes and large-scale wave features (Titov et al., 2012; Peralta et al., 2015), winds at cloud level can be retrieved in UV images and in IR images close to 1 μm (Sánchez-Lavega et al., 2016b). Successful attempts to image Venus surface at its nightside also close to 1 μm are also a current frontier in the capability of amateur astronomers to obtain scientifically valid observations.

- Mars: In spite of the presence of several simultaneous spacecraft orbiting the planet and rovers on Mars surface, global coverage of the planet is still difficult to obtain. Amateur observations of Mars serve to study the global distribution of clouds (especially in short wavelengths, UV to blue, 330-480 nm) onset of dust storms (yellow and green wavelengths, ~500-600 nm), seasonal cycles of the polar caps and albedo changes (red wavelengths, ~600-950 nm). Parker et al. (1995) offer several examples. High-altitude limb cloud-like features discovered in amateur observations is a promising area of research. The last case reported of a very high altitude aerosol layer was singular (Sánchez-Lavega et al., 2015) and its nature is still subject of debate. Future global observations of the planet by amateurs may provide more data on similar events.

- Jupiter: Amateur observations of the planet serve to perform a regular monitoring of the meteorological activity of the planet. This includes: Onset of storms; belt and zones changes; evolution of the Great Red Spot and other large-scale ovals and their interactions. The large amount of high-quality observations in the last year allow also studies of long-lived wave systems (Legarreta et al., 2016). Since at least 2012, amateur astronomers provide images of Jupiter with enough level of detail and regularly enough to measure zonal winds in the planet with an accuracy of 0.1 deg in latitude and a precision of <10 m/s in zonal speeds over several months each year (Barrado-Izagirre et al., 2013; Hueso et al., 2016b). Searches of images with the same central meridian (see Figure 7) allow obtaining excellent amateur image candidates for such an analysis in almost any year since 2010. The overall trend over the last few years in amateur observations of Jupiter has been that there is now a much larger community of well-trained observers able to obtain high-resolution images more often so that a global survey of the atmospheric activity of Jupiter is now substantially better than a few years ago. A large network of amateur observers around the Earth provides now a global coverage of Jupiter when it is close to opposition (Wesley et al. 2009). This might be an essential element in the study of impacts in the planet (Sánchez-Lavega et al., 2010; Hueso et al, 2010b, 2013).

- Saturn: Just like in Jupiter, images of the planet serve to perform a regular monitoring of the meteorological activity of the planet. This includes: onset of storms and large-scale vortices, waves, evolution of the global colors and long-term studies of particular areas like the North Polar hexagon and associated jet (Sánchez-Lavega et al., 2014a). Recently, amateur observations of the planet have contributed to the study of the evolution of the equatorial winds (Sánchez-Lavega et al., 2016a) and polar perturbations in 2014-2016. Global zonal wind profiles are not yet possible to obtain with amateur data, however long-term studies involving different seasons are possible with PVOL data (Figure 8).

- Uranus: Storms, or at least bright discrete features, have been imaged by amateurs allowing to infer their drift rates and plan observations with large facilities (Sromovsky et al, 2012; de Pater et al., 2015).

- Neptune: Since 2013, amateur astronomers have been able to repeatedly observe atmospheric patterns in Neptune (Hueso et al., 2016a). These observations have helped to gather data from several large facilities and simultaneous analysis of amateur observations and data from professional sites is ongoing.

Many of the science topics related with the study of the atmospheres of the Giant Planets have been the subject of different observing programs with the Hubble Space Telescope (HST). However, with the future decommission of the telescope most of the temporal studies of the Giant planets will require significant contributions from amateur astronomers.

## 5. Conclusions

PVOL2 is an advanced database of amateur observations of Solar System planets and the Moon that we hope will be useful for several research teams on different fields and over a long time frame. VO-compliancy provides powerful search functions in the database, the ability to perform cross searches with similar images repositories, and increases the accessibility of global, contextual images in support of more dedicated studies (e.g. using space-borne data). The entry webpage contains news and reports on the current activity on the different planets and of useful observations where amateurs can participate in scientific research. We recognize the important role of amateur astronomers in providing their data and the PVOL2 website includes a full section on publications produced from the data, many of them co-authored by key observers in different research publications (27 publications are listed in the website). PVOL2 also will allow to perform citizen science in a broader context since image tagging can be done by volunteers (identifying all Jupiter images with the Great Red Spot or a given atmospheric feature or projects related with the Moon) on different open calls. Finally, images from PVOL2 can also be a useful tool for teaching atmospheric sciences and related topics at different university levels (Sánchez-Lavega et al., 2014b).

**Acknowledgements**

We thank the continuous support and contributions from all observers contributing to PVOL. They are too numerous to mention all here but we are very thankful to their continuous efforts to observe the solar system providing data of scientific quality. We are particularly grateful to, T. Barry, M. Delcroix, E. Morales, D. Peach, J. Rogers, M. Vedovato and A. Wesley for improvement suggestions, site testing and upload of new data demonstrating new capabilities not present in the previous PVOL database. We thank two anonymous reviewers for their constructive comments. This work has been developed in the framework of the Europlanet 2020 RI. Europlanet 2020 RI has received funding from the European Union's Horizon 2020 research and innovation programme under grant agreement No 654208. This work has also been supported by the Spanish MINECO project AYA2015-65041-P (MINECO/FEDER, UE), Grupos Gobierno Vasco IT-765-13 and UFI11/55 from UPV/EHU.